\documentclass[a4paper,10pt]{article}

\usepackage{graphicx,amsmath,amssymb}
\usepackage{bm}
\usepackage{color}
\usepackage{setspace,textcomp}
\usepackage{authblk}

\begin{document}
 
\title{Stationarity of the inter-event power-law distributions}
\author[1]{Yerali Gandica}
\author[2]{Joao Carvalho}
\author[2]{Fernando Sampaio Dos Aidos}
\author[1]{Renaud Lambiotte}
\author[1]{Timoteo Carletti}
\affil[1]{Department of Mathematics and Namur Center for Complex Systems - naXys, University of Namur, rempart de la Vierge 8, B 5000 Namur, Belgium}
\affil[2]{{CFisUC}, Department of Physics, University of Coimbra, 3004-516 Coimbra, Portugal}

\date{\today}

\maketitle
\singlespacing

\begin{abstract}
A number of human activities exhibit a bursty pattern, namely periods of very high activity that are followed by rest periods. 
Records of these processes generate time series of events whose inter-event times follow a probability distribution that 
displays a fat tail.
The grounds for such phenomenon are not yet clearly understood. In the present work we use the freely
available Wikipedia's editing records to unravel some features of this phenomenon. 
We show that even though the probability to start editing is conditioned by the circadian 24 hour cycle, 
the conditional probability for the time interval between successive edits at a given time of the day is independent from the latter. 
We confirm our findings with the activity of posting on the social network Twitter.
Our result suggests there is an intrinsic humankind scheduling pattern: after overcoming the encumbrance to start an activity, 
there is a robust distribution of new related actions, which does not depend on the time of day.   
\end{abstract}
\newpage \baselineskip1.0cm
\singlespacing
\section{Introduction}

The digital media are an important component of our lives.
Nowadays, digital records of human activity of different sorts are
systematically stored and made accessible for academic research. 
Hence a huge amount of data became available on the past couple of decades,
which allows for a quantitative study of human behaviour.  
For a long time, scholars from different backgrounds have been studying this
field. However, some interesting and basic properties still remained
out of reach for researchers, mainly for lack of large amounts of reliable stored data. 
The increasing amount of data that is being gathered in this digital age is
progressively opening up new possibilities for quantitative studies of these
features.  
One such aspect, detected by means of data-gathering, is human bursty
behaviour, that is the activity characterized by intervals of rapidly
occurring events separated by long periods of inactivity~\cite{barabasi2005}.  
The dynamic of a wide range of systems in nature displays such a behaviour \cite{hidalgo}.

Given the highly non-linear nature of human actions, their study could
hence benefit from the insights provided by the field of complex systems. For
the human being, the bursty behaviour phenomenon has been found to modulate
several activities, such as sending letters, email messages and mobile text messages, as well as making phone calls
and browsing the web
\cite{barabasi2006,barabasi2008,Schellnhuber,amaral2009,kartez2012}.  
The first works in this field suggested a
decision-based queuing process, according to which the next task to be
executed is chosen from a queue with a hierarchy of importance, in order 
to explain the observed behaviour. 
Different kinds of hierarchies were tested, such as the
task length and deadline constraints
\cite{barabasi2005,barabasi2006,barabasi2008}. Later on, Malmgren
{\it et al.} \cite{amaral2008,amaral2009} argued that decision making is not a
necessary component of the bursty human activity patterns. Instead, 
they maintained that this feature is caused by cyclic constraints in life and
they proposed a mechanism based on the coupling of a cascading activity to
cyclic repetition in order to explain it. Nonetheless, recently, Hang-Hyun Jo
{\it et al.} \cite{kartez2012} applied a de-seasoning method to remove the
circadian cycle and weekly patterns from the time series, and obtained similar 
inter-event distributions, before and after this filtering procedure. 
In this way, the authors concluded that cyclic activity is also not a necessary ingredient of bursty behaviour. 

The goal of the present work is to contribute to the issue of human burstiness universality, by looking at Wikipedia editing and
Twitter posting. In particular, we show that the same inter-event distribution happens at each hour 
of the day. We relate this kind of universality, the result of a single person's decision, to a kind
of resource allocation (attention, time, energy), distributed in proportion to the different activities that the
individual is able to do at specific times, and which is responsible for the broad distribution of inter-events, characteristic of
a bursty behaviour. The bursty nature independence on the high or low activity, as a result of circadian patterns, 
is an important issue when trying to predict human activity in social media platforms \cite{harada,kobayashi,leskovec}.

\section{Methods}
Our study explores the editing activity of the super-editors (defined hereafter) in four separate Wikipedias (WP) \cite{data}, written in four different
languages: English (EN-WP), Spanish (ES-WP), French (FR-WP) and Portuguese (PT-WP). 
In all cases the data span a period of about ten years, ending between 2010 and 2011 (depending on the language). 
Each entry in the database contains the edited WP page name, the time stamp of the saving and the identification of the 
editor who performed the changes. Moreover, we discarded entries associated to IPs and 
bots, and only considered editors who login before editing, so that the editor is univocally identified.

\begin{figure}[tbh]
\begin{center}
\scalebox{0.34}{\includegraphics{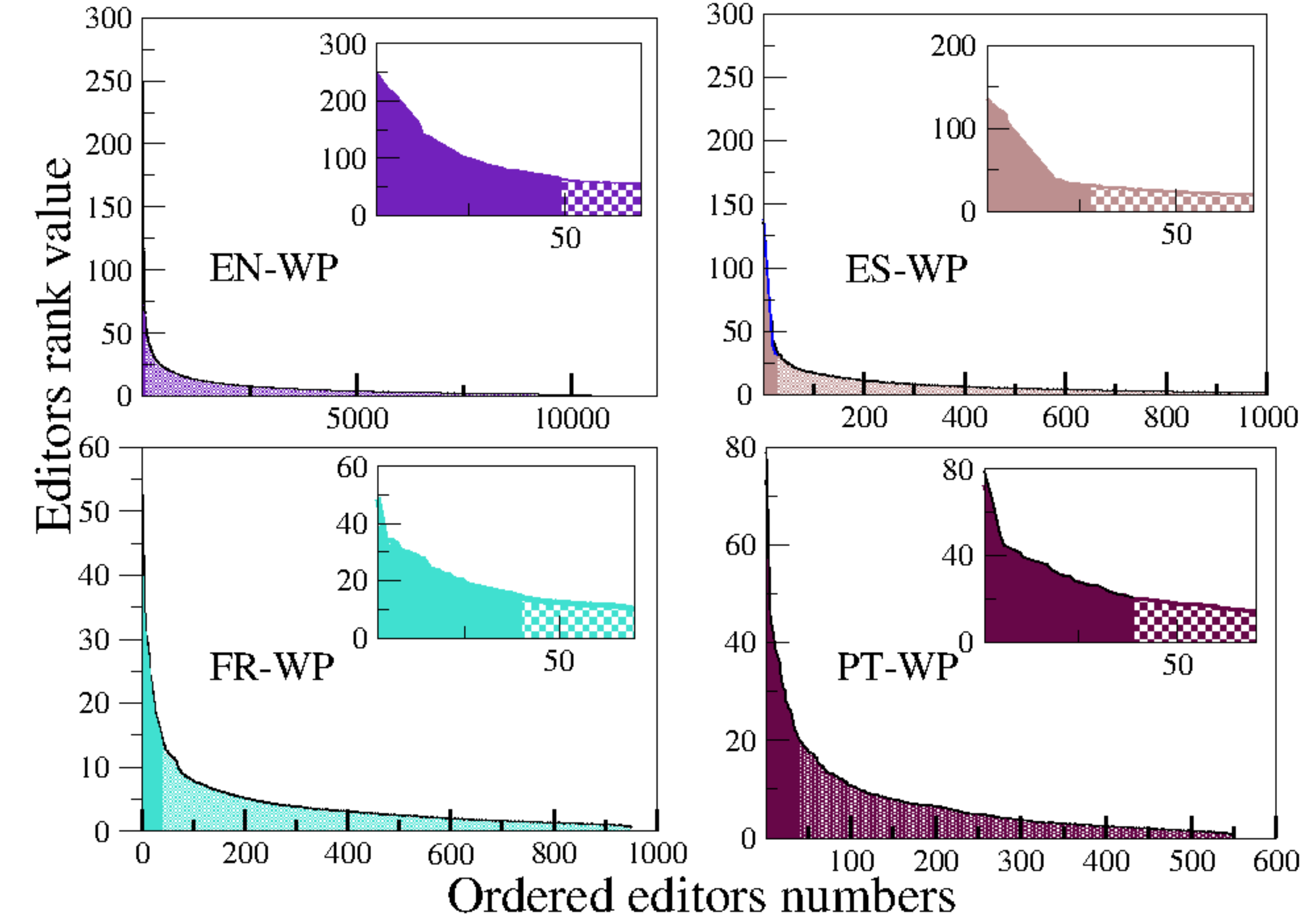}} \caption{{\bf Normalized activity for all 
the editors with more than $2000$ edits, for the four WP's}. The darker areas show the super-editors zone. 
The inset in each figure displays a zoom for a better visualization of the super-editors region. 
Some editors in the darker zone were not considered super-editors because they did not edit for more than one year.} 
\label{fig1}
\end{center}
\end{figure}

Only editors with more than $2000$ edits are considered, in order to reduce the impact of outliers and to have enough active editors in the data set. 
After the filters, the universe of our sample is composed by $10473$ editors in EN-WP, $1110$ in
ES-WP, $955$ in FR-WP and $551$ in PT-WP. 
We define the normalized activity, or rank, of an editor as his total number of edits divided by the 
total number of days since he started to edit. 
Super-editors, in a given language, are defined as the editors whose normalized activity is greater than $25 \%$ of the largest normalized activity
in that particular WP and with more than one year of editing activity. 
The number of super-editors is $20$ in EN-WP, $10$ in ES-WP, $15$ in FR-WP and $24$ in PT-WP. 
We have checked that neither WP-bots nor blocked editors are among the super-editors in our list. 
In Fig.~\ref{fig1} we plot the normalized activity for all the editors with more than $2000$ edits in decreasing order, for the four WP's. 
The darker areas in the plots show the regions where the super-editors lie. In each figure, we include an inset that contains a zoom with a better view of
the super-editors zone. 
Note that we have not applied the one-year of activity filter yet, so some 
editors in the darker zone were not considered super-editors. 
We focused on super-editors because their high activity provides suitable statistics; 
moreover, as recently shown~\cite{icwsm}, their behaviour is quite similar to standard editors with respect to the memory coefficient 
$M$ and the burstiness parameter $B$, as defined in~\cite{barabasi2008}.

\section{Results}
In \cite{complenet} we have shown that WP editing is strongly influenced by the circadian cycle, as reported before by Yasseri et al. \cite{kartez2012b}.
Here we analyze whether these circadian patterns have consequences on the inter-event probability distribution, namely we check whether the time between 
edits depends on the hour of the day at which the first edit has been carried out. 
To perform such an analysis we computed the probability 
distribution for the inter-event duration, considering that the first event has taken place at a specific hour of the day. 
If this conditional probability depends on the hour of the day then we can conclude that circadian cycles have an influence 
on the human inter-event time and thus the origin of burstiness can possibly be ascribed to this dependence. 
In the opposite case we can conclude that burstiness in WP editing does not depend on the periodically changing conditions.

\begin{figure}[tbh]
\begin{center}
\scalebox{0.34}{\includegraphics{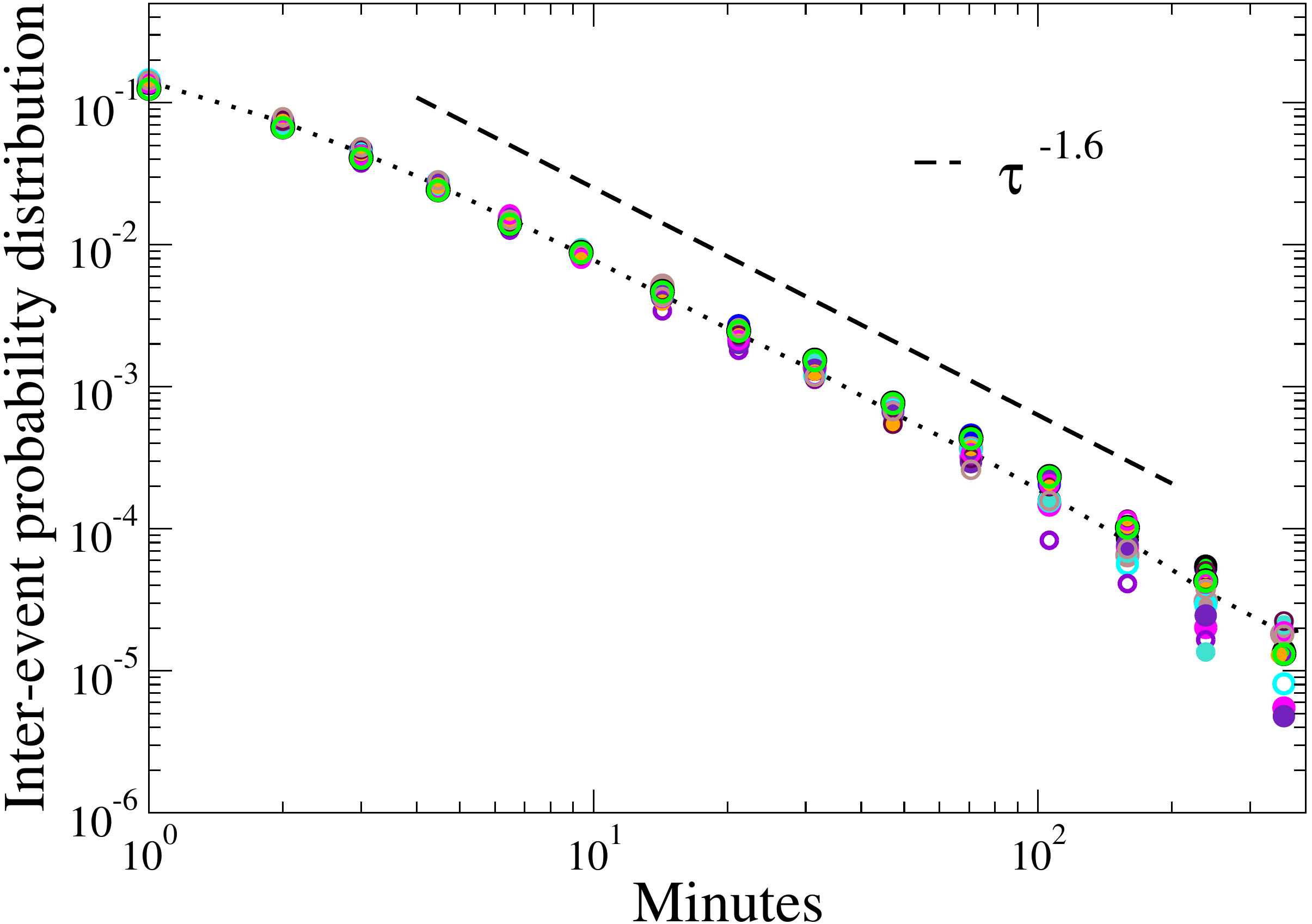} \includegraphics{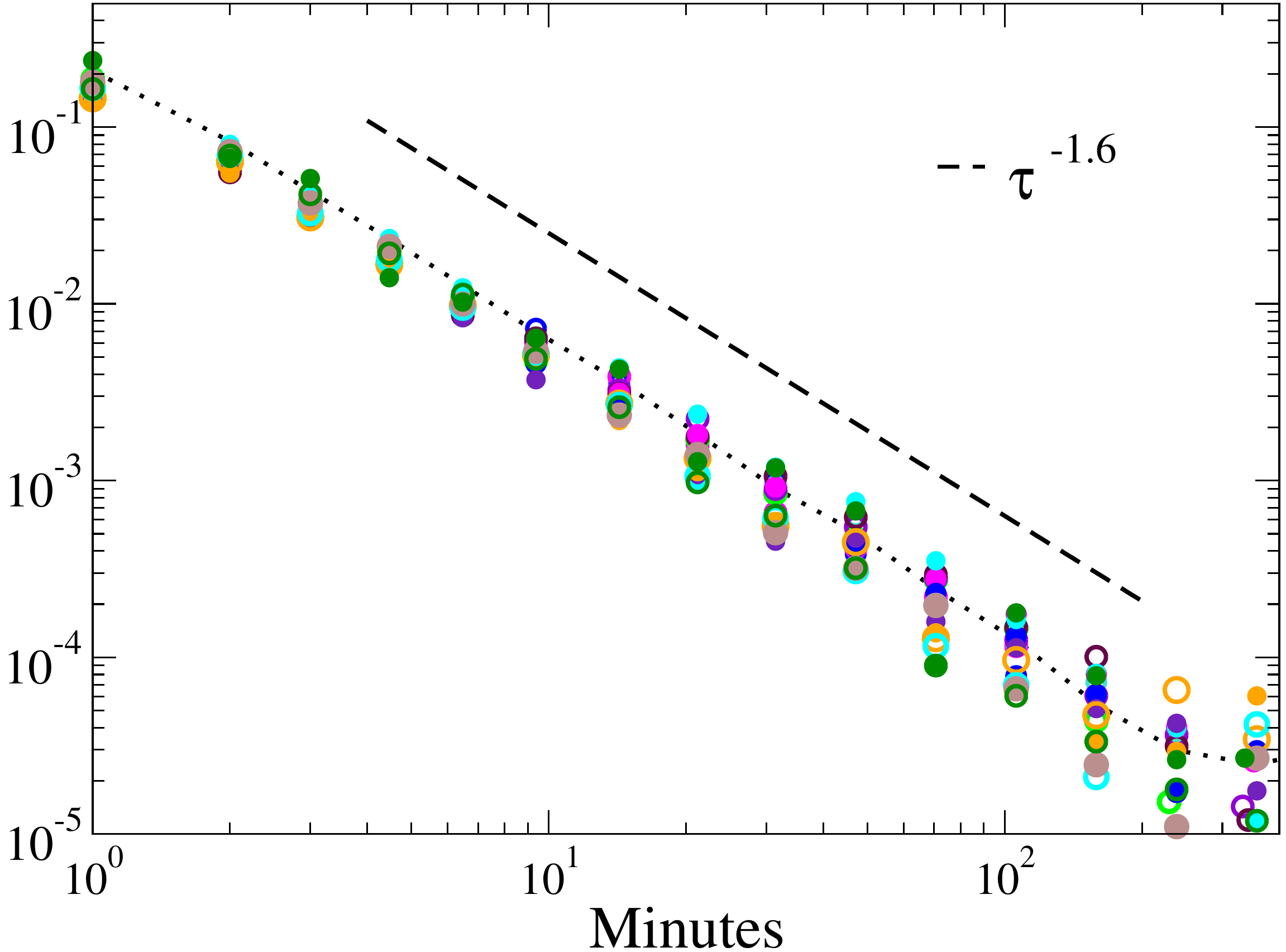}} \caption{{\bf Conditional probability distributions for the inter-event duration.} 
We represent, for two of the most active editors, the conditional probability distribution to have an inter-event activity of duration $\tau$ given 
the hour of the day at which such action is registered, represented by a different color-symbol for the $17$ one-hour windows used. Seven 
windows were left out because the small quantity of data they contain was not sufficient to draw statistically 
sound conclusions. 
The dotted lines represent the inter-event probability distribution using a window of $24$ hours, containing all the data. 
One can clearly note the similar fat-tails in all the time windows, indicating they are independent from the circadian cycle. 
In both panels, the dashed lines represent the power law best fits, whose exponents are $-1.59 \pm 0.04$ for the editor shown in the left panel 
and $-1.58 \pm  0.09$ for the editor in the right one.} 
\label{interdiarydistri}
\end{center}
\end{figure}

Results reported in Fig.~\ref{interdiarydistri} support the latter hypothesis. 
The inter-event conditional probability distributions computed in different one-hour
windows -- large enough to contain adequate statistics but not too large to avoid averaging features -- exhibit a similar fat tail 
when they are normalized by the number of events in that time window. 
Note that only $17$, out of $24$, time windows are shown in Fig.~\ref{interdiarydistri}; seven windows are associated to low activity 
periods, and data are scarce in these windows. Therefore, these windows were discarded as they could introduce spurious effects 
(see also Fig.2 in~\cite{complenet}). We also limited the maximum inter-event duration to seven hours, once again to avoid spurious effects in the queue distribution.
Our fits were done using the software ROOT \cite{root} and compared with the procedure by Clauset {\it et al.}\cite{clauset}. 

Our results seem to indicate that, although the probability of editing is strongly influenced by circadian rhythms, the 
probability distribution for the time between successive edits is indeed independent from them. This suggests that the bursty nature of the process is independent from the circadian patterns. 
Note that a similar result, but on longer time scales, has been previously presented in~\cite{alain2010}, where the
authors reported the robustness of the inter-event time distributions using $12$ hours windows for binary contacts between conference participants.

The use of one-hour time windows is, in our opinion, a good proxy to demonstrate the stationarity of the inter-event distribution during the day. 
One should use even smaller time windows, but this would require a huge data sample to have enough statistics in each small period of time. The
conditional probability to continue an action has been previously simulated by means of cascades of events, triggered by the initial event, which is
conditioned by circadian patterns, by Malmgren {\it et al.} in \cite{amaral2008}. 

\begin{figure}[tbh]
\begin{center}
\scalebox{0.34}{\includegraphics{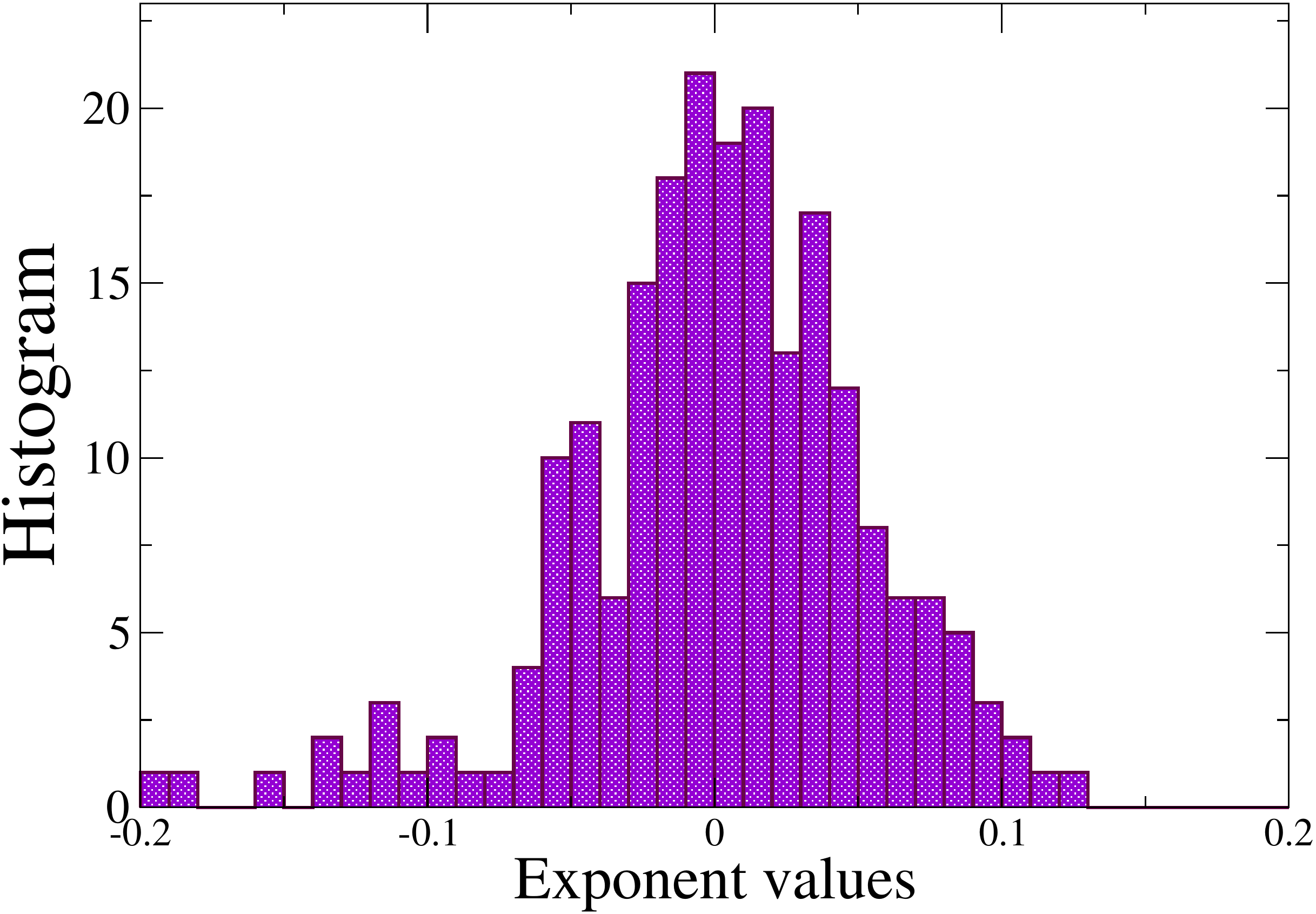}} \caption{{\bf Distribution of the power law exponent relative deviation}. 
$(\alpha^{(j)}_i - \langle \alpha^{(j)} \rangle) / \langle \alpha^{(j)} \rangle$, for all super-editors 
and time windows with adequate statistics, in the EN-WP.} 
\label{fig3}
\end{center}
\end{figure}

The fat-tail distributions presented in Fig.~\ref{interdiarydistri} can be well described by a power law $p(\tau | t)=c\tau^{\alpha}$, 
where $t$ is the hour of the day when an event takes place and $\tau$ is the time to the next event to take place. 
The exponent $\alpha<0$ is independent of the hour of the day $t$. In order to study the variability of the exponent values for different
super-editors and time windows, we fit, for each super-editor, $j$, and each time window, $i$, the data in Fig.~\ref{interdiarydistri} 
and obtain the power law exponent $\alpha_i^{(j)}$. 
We hence obtain the average exponent $\langle \alpha^{(j)} \rangle=\sum_i \alpha_i^{(j)}/N_{j}$, being $N_j$ the number of
windows used for super-editor $j$, and finally we compute the relative deviation 
$(\alpha^{(j)}_i - \langle \alpha^{(j)} \rangle) / \langle \alpha^{(j)} \rangle$. 
A histogram with the probability distribution of the relative deviation for the EN-WP is shown in Fig.~\ref{fig3}. The behaviour is
well described, apart from some fluctuations, by a normal distribution with standard deviation $0.055$.


\begin{figure}[tbh]
\begin{center}
\scalebox{0.8}{\includegraphics{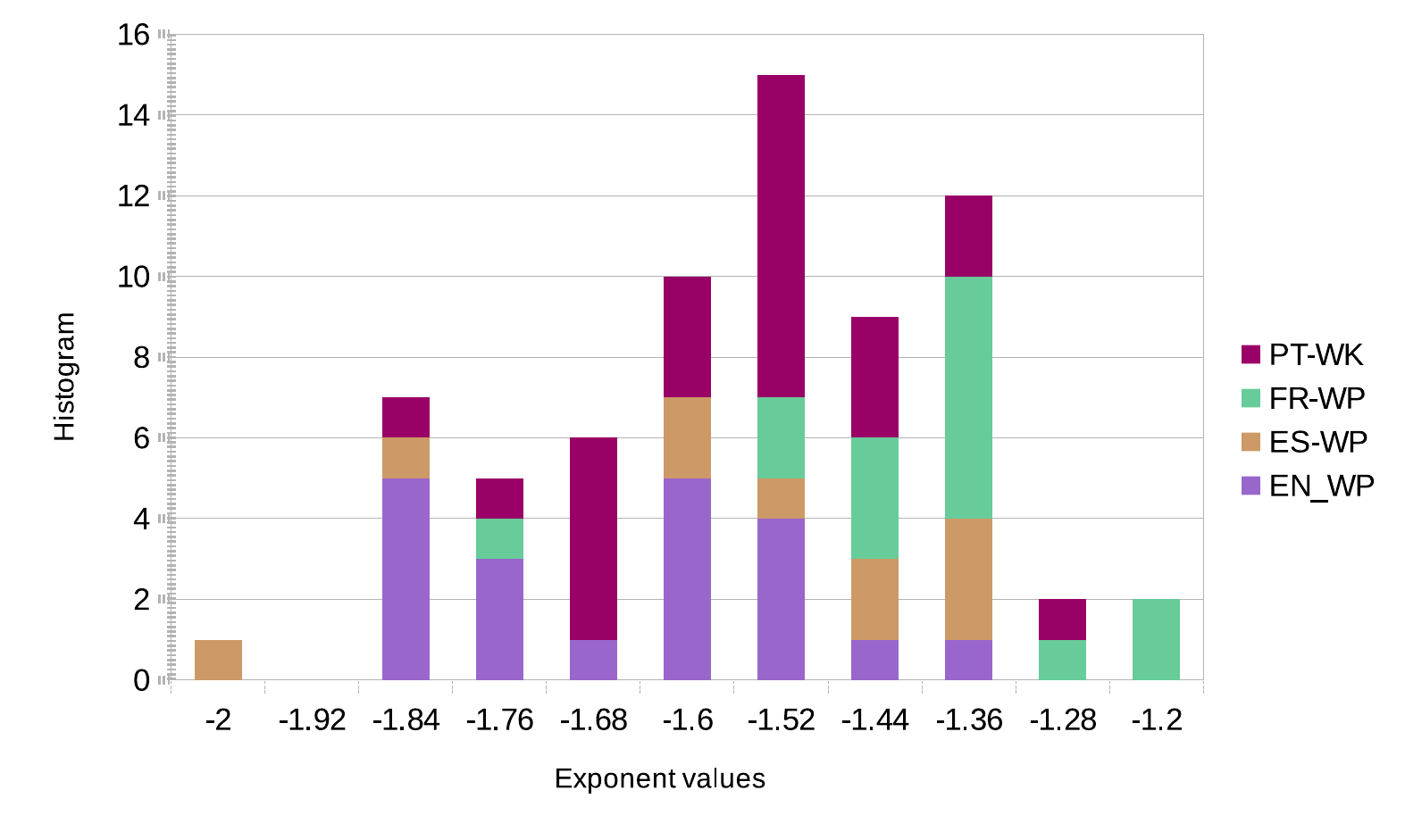}} \caption{{\bf Distribution of the averaged exponents $\langle \alpha^{(j)} \rangle$ for the super-users in the 
four WP's.}} 
\label{fig4}
\end{center}
\end{figure}

We report the distribution of the average exponents for all the inter-events, $\langle\alpha^{(j)}\rangle$, for the super-editors in the four WP's in Fig.~\ref{fig4}. 
We notice that the average value is $-1.59$; note that in~\cite{kartez2013} the average value of the 
exponents computed using $24$ hours long windows for the $100$ most active WP editors, was reported as $-1.44$. 

\begin{figure}[tbh]
\begin{center}
\scalebox{0.3}{\includegraphics{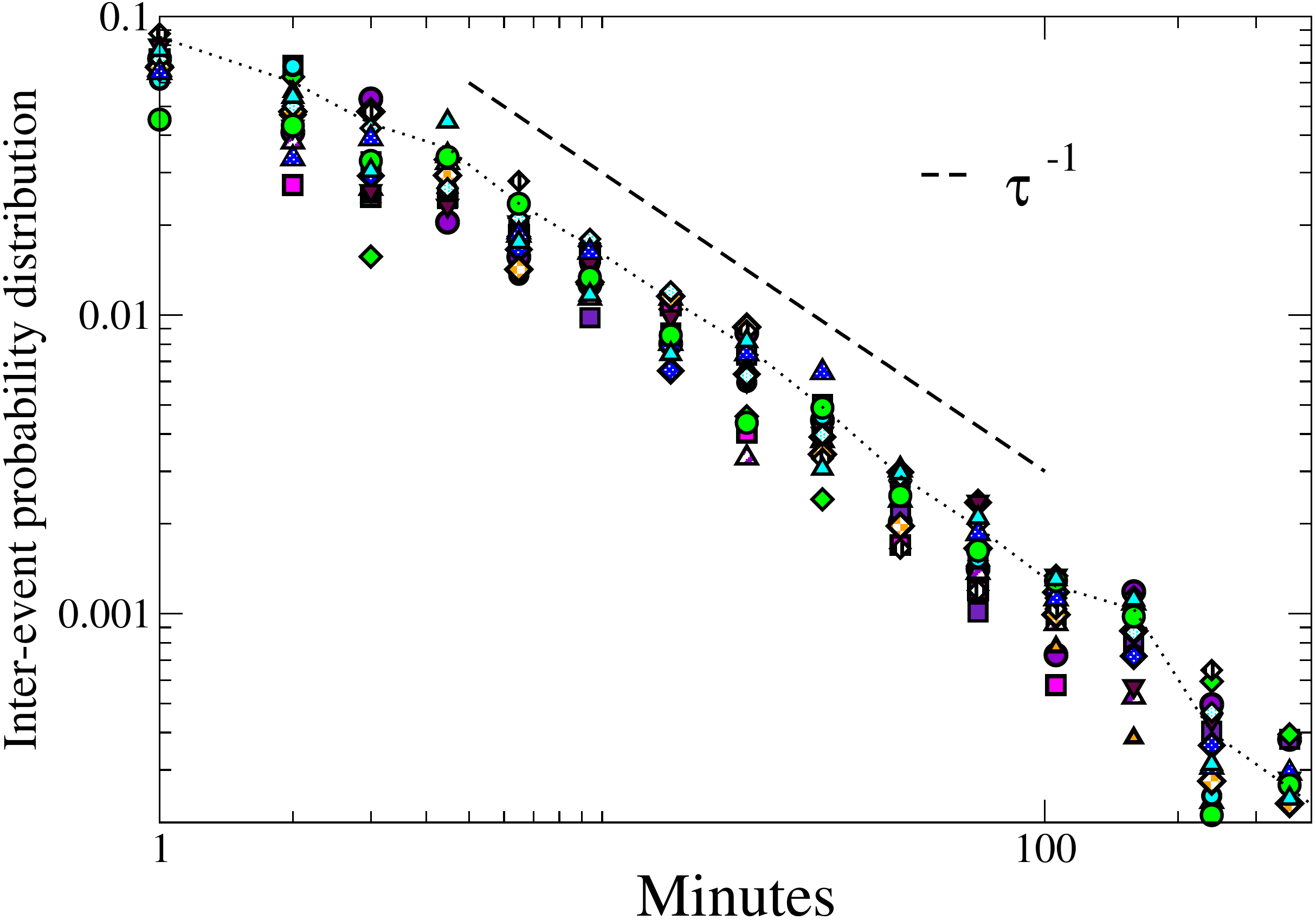} \includegraphics{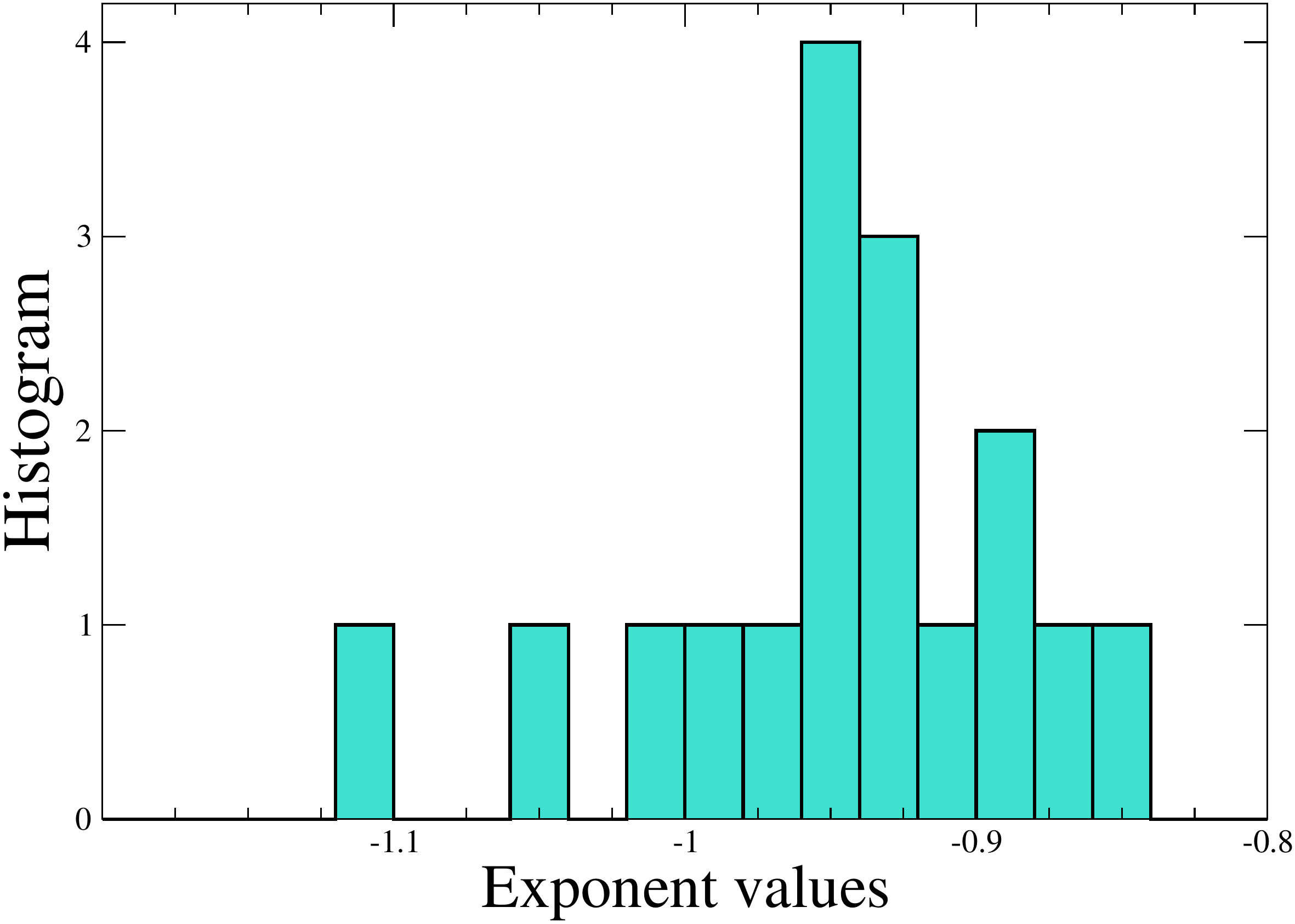}} \caption{ {\bf Left panel: Conditional probability 
distributions for the inter-event duration of all the tweets posted by a user who started in February 2010, with a total
of $8266$ tweets. Right panel: distribution of the power law exponents from a fit to each of the 17 hourly inter-event distributions.}
Each color-symbol in the left panel represents the probability distribution of all the inter-events registered in each one-hour window. 
Seven windows were left out because the small amount of data they contain was not statistically relevant. The dotted lines 
represent the inter-event probability using a window of $24$ hours, containing all the data. The dashed line represents the 
power law exponent best fit $-0.96 \pm 0.07$.} 
\label{fig5}
\end{center}
\end{figure}

In the formulation of several hypothesis trying to understand the origin of the robustness of the power-law distribution
among the several hours of the day, we started with the 
hypothesis of the saving mechanism, the time between successive savings when editing the same page, something that could be 
of behavioral origin across human activity. This hypothesis can perfectly explain the independence of the inter-event 
distribution with respect to the time of the day. However, we looked into the pages edited by the same editor and short 
inter-events were found regularly also between different pages editing, and in consequence they cannot be a simple act of saving the work done so far in a certain WP page.  
This could be enough to discard the saving process mechanism as the origin, but the low probability that the editor has opened 
several windows to edit in parallel is still present. 
Because we are interested in the mechanism as the result
of an internal process inherent to the general human activity, we complement the analysis with the activity of posting in 
a social micro-blogging platform, Twitter. Each posted tweet consists of a message of $140$ characters. 
In the left panel of figure \ref{fig5}, we show the inter-event conditional probability distributions computed in different 
one-hour windows, for all the tweets posted by one representative user, starting in February 2010. 
Following the same procedure as before, we discard the hourly intervals of low activity (seven intervals, in this case), and 
we show the inter-event up to 7 hours. Again, we found the same power-law distribution for the inter-event times across the 
different hours of the day, but now with an exponent of about -1. 
In the right panel of the same figure the distribution of the 
power-law exponent best fit values in each active hour of the day is depicted.

\section{Discussion}
To summarize, in this work we provide numerical evidence that the conditional probability, $p(\tau|t)$, to have an inter-event of duration $\tau$ for 
an edit of WP registered at time $t$, is independent from the latter. Moreover this probability is fat-tailed, well described by a power law. 
It could be related to some sort of queuing process, but we prefer to see it as due to a resource allocation (attention, time, energy) process, 
which exhibits a broad distribution: shorter activities are more likely to be executed next than the longer ones,
which ultimately may be responsible for the bursty nature of human behaviour. 

Using the data for the editing of WP and for the activity of tweeting, our results seem to indicate that there is an 
intrinsic mechanism to human nature: before performing an action (make a phone call, send a tweet, edit 
Wikipedia, etc) we must overcome a ``barrier", acting as a cost, which depends, among many other things, on the time of day. However, once that ``barrier" 
has been crossed, there exist a robust distribution of activities, which no longer depends on the time of day at which we decide to start it.
Our findings suggest that the bursty nature of human beings is mainly independent of circadian patterns, in agreement with the
results found, using a different method, by Hang-Hyun {\em et al.} \cite{kartez2012}. This result could open the perspective to 
applications less specific than the study of Wikipedia. Future work includes simulations taking into account circadian
patterns to reproduce the probability to perform an action, while maintaining a constant conditional probability distribution for 

\vspace{1cm}
\textbf{Acknowledgments}
The work of Y.G., T.C. and R. L. presents research results of the Belgian
Network DYSCO (Dynamical Systems, Control, and Optimisation), funded by the
Interuniversity Attraction Poles Programme, initiated by the Belgian State, Science Policy Office.

\end{document}